# Modulated 3D cross-correlation light scattering: improving turbid sample characterization

Ian D. Block and Frank Scheffold

Department of Physics, University of Fribourg, Fribourg, Switzerland

LS Instruments AG, Rte de la Fonderie 2, Fribourg, Switzerland

ian.block@lsinstruments.ch
frank.scheffold@unifr.ch

## Abstract

Accurate characterization using static light scattering (SLS) and dynamic light scattering (DLS) methods mandates the measurement and analysis of singly-scattered light. In turbid samples, the suppression of multiple scattering is therefore required to obtain meaningful results. One powerful technique for achieving this, known as 3D cross-correlation, uses two simultaneous light scattering experiments performed at the same scattering vector on the same sample volume in order to extract only the single scattering information common to both. Here we present a significant improvement to this method in which the two scattering experiments are temporally separated by modulating the incident laser beams and gating the detector outputs at frequencies exceeding the timescale of the system dynamics. This robust modulation scheme eliminates cross-talk between the two beam-detector pairs and leads to a four-fold improvement in the cross-correlation intercept. We measure the dynamic and angular-dependent scattering intensity of turbid colloidal suspensions and exploit the improved signal quality of the modulated 3D cross-correlation DLS and SLS techniques.

## Introduction

Static and dynamic light scattering are powerful and widely employed optical characterization techniques that rely on the interaction of coherent light with density fluctuations in a small volume [1-4]. The source of these density fluctuations can for example be polymers in solution, small particles in suspension, surfactant micelles, or subcellular biological components. When such a sample is illuminated with a laser source, the scattered light yields a speckle pattern in the far field. The spatial intensity distribution of speckles is dictated by the summation of the angular-dependant scattering efficiencies of the density fluctuations in the illumination/detection volume and by the phase relationship of the scattered fields. For particles in suspension, these are determined by particle shape, size and their relative positions. Static light scattering (SLS) involves the measurement of angular-dependent time-averaged scattering intensities, and provides precise information on size and shape as well as the structure of interacting samples[1]. Again using the example of a particle suspension, as particles displace randomly in time due to Brownian motion, the intensity of the speckles fluctuate as the phase relationship of the light scattered by the particles change. Dynamic light scattering (DLS) refers to the measurement of the intensity fluctuations of a single speckle and the subsequent computation of a correlation function. The decay rate of this function can lead directly to the extraction of diffusion coefficients and to the determination of size distributions through straightforward application of the Stokes-Einstein equation. This method has become a standard for particle size analysis in the range of a few nanometers to a few microns [4, 5].

Both static and dynamic methods rely on the measurement of single scattering events, meaning that each detected photon has been scattered only once in the sample[6-8]. Therefore, dilution is typically a necessity for highly scattering and concentrated samples. However, the dilution process and especially the verification of adequate dilution can be extremely time-consuming. Furthermore, for samples wherein a concentration-dependent behaviour is of interest, dilution is not an option. Index matching of the solvent can be accomplished in some cases, but this in general has limited applicability to a few model systems [9]. Narrow sample confinement can be effective in minimizing multiple scattering[11]. Additionally reducing the scattering volume using a microscope-based approach maintaining well-defined scattering vectors was also shown to be capable of measuring turbid samples[12]. However, these methods suffer from the drawback of a limited range of accessible scattering vectors. They also require extremely small sample volumes that are difficult to handle and may be subject to drying or contamination during handling. Thus attempts to minimize multiple scattering by reducing sample and scattering volume only have limited applicability. Another related technique that can be applied to larger sample volumes employs the use of a single-mode fiber to deliver and collect laser light [13]. The acceptance cone of the fiber is narrow and will exclude multiply scattered light if the mean free path in the turbid sample is large enough. However, the geometry of this detection method limits its utility due to an inability to measure an angular-dependent response.

There are other techniques that instead seek to suppress the influence of multiple scattering in the optical signal, and employ a more standard goniometer-based measurement setup which provide a large range of measurable scattering vectors. These techniques rely on the cross-correlation of two measurements to extract single-scattering information from the same scattering volume and the same nominal scattering vector. The first of these methods uses a single incident laser beam and two detectors which are separated by the characteristic length scale of a single speckle [14-16]. Suppression is achieved because multiple scattering occurs over a larger volume in space (as determined by the mean free path) and therefore gives rise to speckles in the far field smaller than the detector separation distance. However, this method typically has a cross-correlation signal amplitude significantly below that of the ideal case of one, and the signal degrades rapidly with moderately turbid systems. The remainder of the methods also involves the cross-correlation of two measurements performed on the same scattering volume, but uses two illumination beams aligned such that a degenerate pair of scattering vectors is measured[7, 8, 17]. Single-scattering information is then common to both measurements while multiple scattering information is uncorrelated, thereby leading to its effective suppression. Although developed to suppress multiple scattering for DLS, these methods have also been shown to be effective for SLS measurements of turbid samples

Multiple scattering suppression can be accomplished in practice by using two lasers operating at different wavelengths and two detectors having distinct bandpass filters to capture scattering information from each single laser. This technique is generally referred to as two-color DLS[18]. While the two-color technique has proven to be effective at extracting single-scattering information from highly turbid systems, the technical challenges of obtaining and maintaining precise alignment of the illumination and detection optics, especially as the scattering vector is varied, makes implementation and operation all but prohibitive[19]. Although in the 1990's a commercial product was marketed, to our knowledge it has been discontinued mainly for this reason.

Another more simple and robust arrangement of the beam-detector pairs is possible by displacing them symmetrically in a third dimension, where this is known as the 3D cross-correlation geometry [7, 10, 17, 20, 21]. A commercial goniometer setup is available from LS Instruments AG (Fribourg, Switzerland) since 2001 based on the 3D arrangement and is now routinely used by a number of laboratories around the world. However, one important drawback of the 3D technique compared to two-color DLS is that one photon detector measures the scattered light intensity at the desired scattering vector, but also receives a contribution at a second undesired scattering vector given by the relative geometry to the second illumination beam operating at the same wavelength[19]. A four-fold reduction in the cross-correlation intercept arises from cross-talk between the two simultaneous scattering experiments executed in this way. The intercept refers to $\beta$, the adjusted $y$-intercept of the intensity cross-correlation function given by

$$g_2(q,t) = \frac{\langle I(q,0)I(q,t)\rangle_T}{\langle I(0)\rangle_T^2} = 1 + \beta|g_1(q,t)|^2 \quad (1)$$

where $I$ is the measured intensity at a given scattering vector $q$ and time $t$, brackets indicate an ensemble average taken over time $T$, and $g_1$ is the normalized field correlation function. The value of $\beta$ strongly influences measurement accuracy and precision due to its pivotal role in accurately fitting models to the measured data. For strongly scattering samples where only a small component of the detected light is singly-scattered, the signal-to-noise ratio of the measurement becomes unacceptably low as the magnitude of the cross-correlation intercept falls into the noise of the baseline fluctuations.

One method that attempts to achieve complete signal separation in 3D cross correlation involves the use of orthogonally polarized incident beams in addition to polarizers placed in front of each detector[22]. However, application of this approach is limited to spherical particles and a small range of forward scattering angles where scattering is insensitive to the polarization state of incident light.

In order to realize a practical light scattering instrument capable of making complete DLS and SLS measurements in highly scattering samples, we present here a 3D cross-correlation light scattering instrument in which the two beam-detector pairs are temporally isolated. The illumination beams are alternately activated with high speed intensity modulators and the detection electronics are gated in unison. A schematic as well as a photograph of the instrument are shown in Fig. 1. We demonstrate a four-fold improvement in the cross-correlation intercept measured relative to a standard 3D arrangement and provide results of DLS and SLS measurements that illustrate significant improvements in data quality.

**Experimental Setup**

Temporally separating the two scattering experiments performed in the 3D geometry mandates precise temporal control of each laser beam and photon detector output. The dynamics of a particular sample of interest dictate the minimum necessary time scale for which cross-correlation data must be collected. In turn, this time scale determines the technology that must be implemented to achieve efficient signal separation. In order to ensure applicability to a wide range of multiply scattering samples, a design goal in the present work was to ensure the ability to capture data at time scales down to a few microseconds. The choice of this limiting time scale is dictated by the Brownian dynamics of small objects. For an estimate we consider small particles suspended in water at room temperature and a laser wavelength of 633 nm. Multiple scattering typically becomes an issue for colloidal systems with size features of diameter 100 nm and above. For much smaller objects the scattering cross section decreases rapidly and samples appear weakly scattering even at elevated densities. A lower bound of the Brownian relaxation time can thus be estimated as $\tau > (4D_B k^2)^{-1} = 0.3\text{ms}$, where $D_B$ is the Brownian diffusion coefficient and $k$ the wavenumber in water. A time resolution of several microseconds should thus be more than sufficient for all situations of practical interest. Therefore, the system has been designed for robust operation at modulation frequencies approaching 1 MHz. Alternate shuttering of the laser beams is accomplished using acousto-optic modulators (AOMs; available through LS Instruments AG) based on tellurium dioxide, having a 2.0mm clear-aperture, and exhibiting a rise-time of approximately 250 ns. The modulators are driven by digitally modulated RF amplifiers that output a 110 MHz sinusoidal driving signal at up to 2W. Each AOM can be oriented such that when it is not activated, the laser beam passes directly through with negligible loss and the optical alignment of the standard 3D cross-correlation instrument is preserved. However, when driven with a high frequency RF signal, a phase grating is formed in the acousto-optic crystal and the incident laser beam is deflected into the first order with high efficiency (>95%). While the deflection angle is very small, adequate spacing between the AOM and the entrance of the scattering cell enables placement of a beam block to prevent passage of the deflected beam. In this fashion, activation of the AOM leads to shuttering of the laser beam. Alternate shuttering of the two beams is performed by using two AOMs, where each device is driven by an independent RF amplifier that is controlled by a digitally modulated signal, in which the signal for one amplifier is the logical inversion of the other.

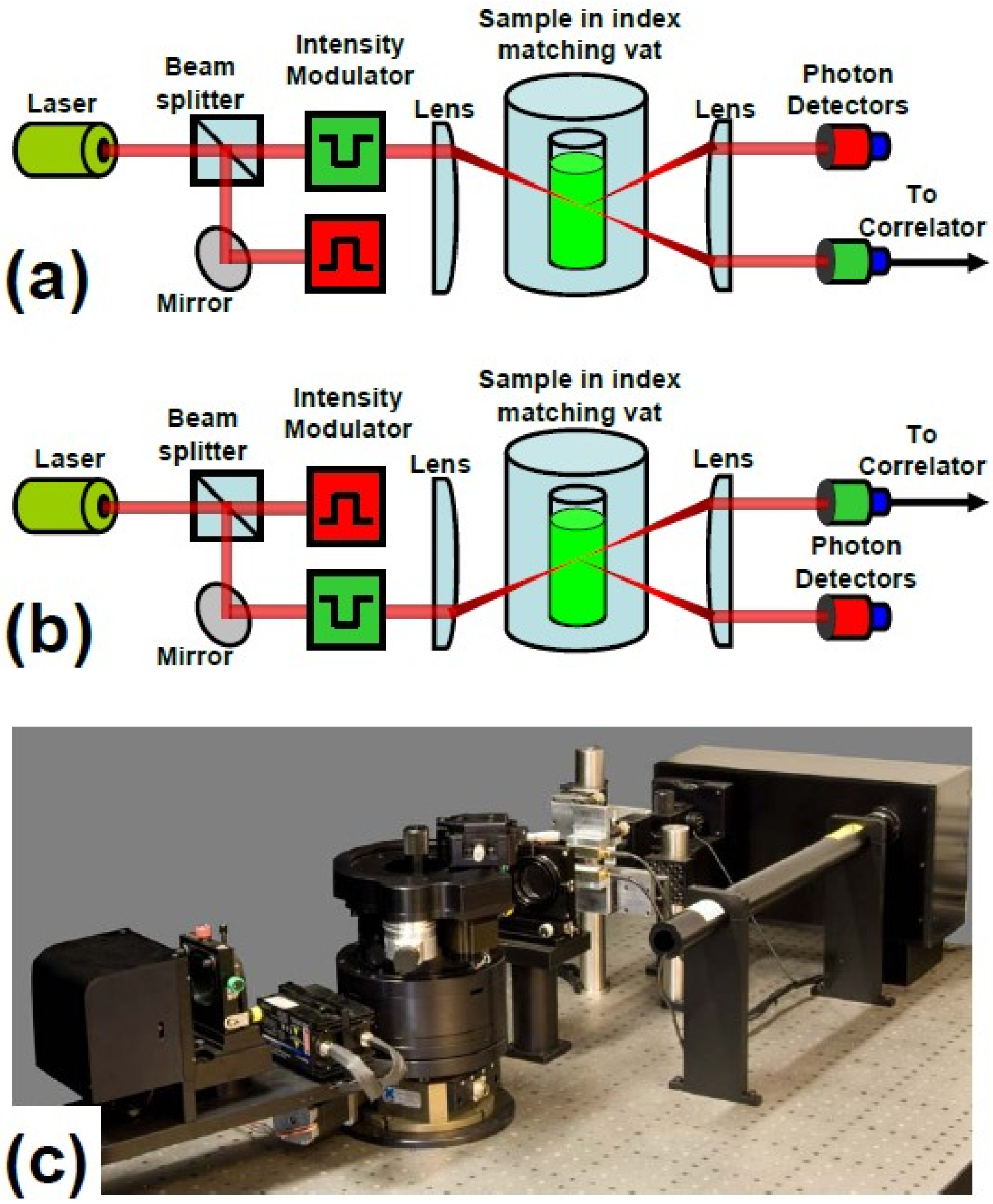

**FIG. 1.** (Color Online) **(a, b)** Schematic of modulated 3D cross-correlation light scattering instrument showing the two states wherein one of the modulators is activated and one detector is gated. **(c)** Photo of modulated cross-correlation setup adapted to a commercial 3D DLS instrument from LS Instruments.

In addition to modulation of the incident beams, the outputs of each photon detector must also be gated in order to assure that only scattered light at the appropriate geometry is recorded for each beam-detector pair. The photon detectors used in this work are avalanche photodiodes (APDs; Perkin Elmer) operating in Geiger mode and generating 15 ns TTL pulses for each incident photon. The simplest gating scheme involves a logical AND of the TTL pulses and the inversion of the digital signal driving the AOM for the incident beam yielding the desired scattering geometry. In this case, when the AOM acting as a shutter is activated, the output of the corresponding APD is effectively 'blanked'; and when the AOM is off such that the laser beam passes into the scattering cell, the digital signal from the APD proceeds as usual to the subsequent hardware. However, the situation becomes more complex as the modulation frequency is increased since the rise and fall time of the AOM response become appreciable relative to the modulation period. In order to avoid the situation in which both beams are simultaneously partially on and therefore there is a degradation of the cross-correlation intercept, the output of the active APD is blanked during the transition time of the AOMs. In this fashion only scattered photon events measured during which the incident laser intensity is constant are passed on to the correlation hardware. This is implemented by similarly performing a logical AND of the output signal from the APD with the digital signal driving the AOM, but additionally modifying the duty cycle of the modulated reference signal in order to control precisely when the detector is blanked. A signal processing board was fabricated to contain elements to generate a variable driving frequency (10 kHz – 1 MHz), control the duty cycle of the blanking signal (with a precision of 10 ns), and RF relays to enable the possibility to bypass the gating electronics. A hardware correlator performs a cross-correlation of the two processed outputs of the APDs, where the minimum lag time is set to coincide with the shutter modulation period. As such, the modulated signals are under-sampled and the resulting cross-correlation function is free of any oscillatory components. All optical and electronic components mentioned above were integrated with a commercial goniometer-based 3D light scattering instrument (LS Instruments AG, Fribourg, Switzerland) such that measurements could easily be made with and without laser beam and detector modulation, and all other instrument functionality was retained.

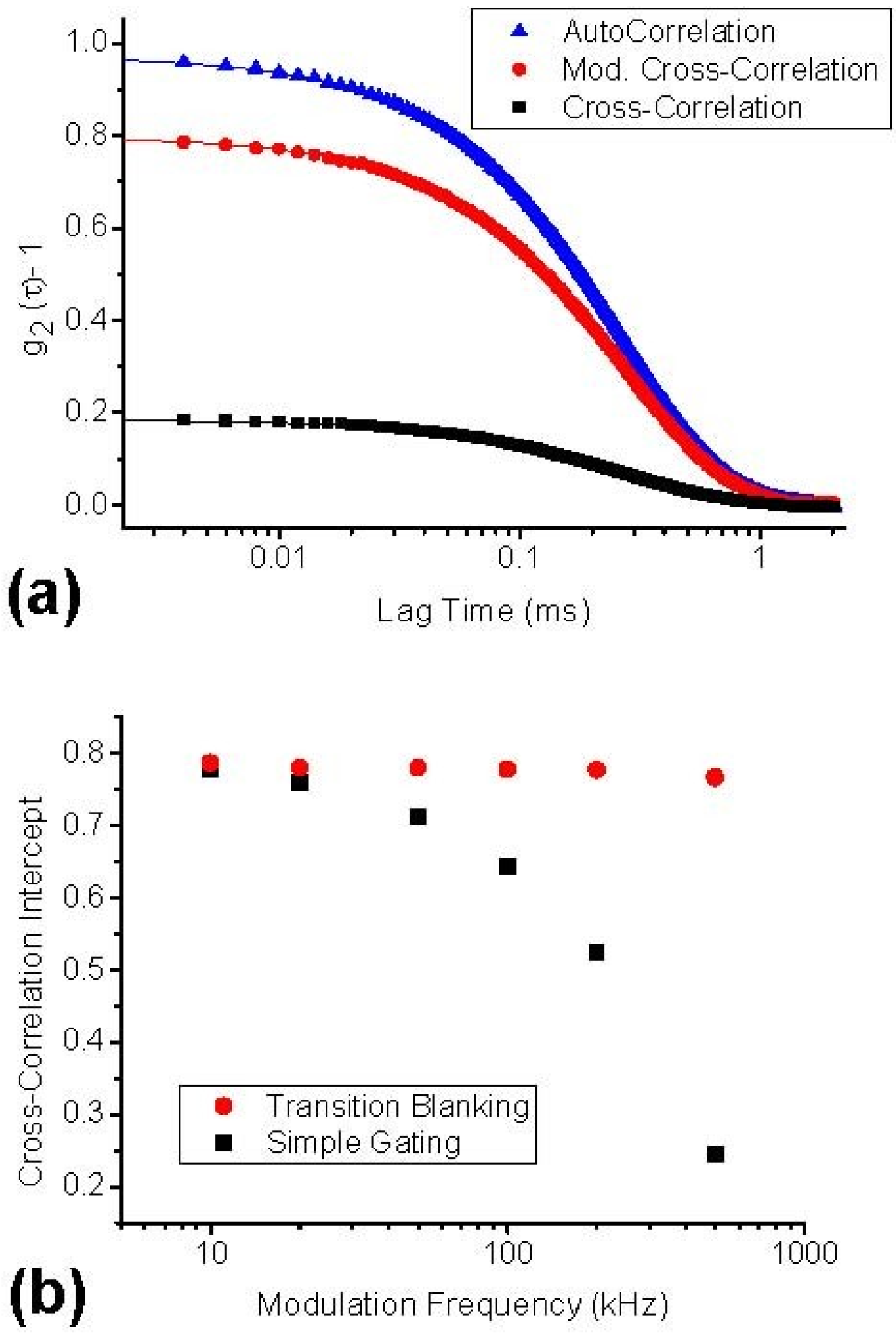

**FIG. 2.** (Color Online) **(a)** Correlation function comparison demonstrating four-fold improvement of the intercept for modulated 3D cross-correlation versus standard 3D cross-correlation. **(b)** Frequency dependence of modulated cross-correlation intercept illustrating necessity at higher frequencies of photon detector blanking during the transition time of the intensity modulators.

**Methods**

In order to characterize the modulated cross-correlation DLS setup, measurements were made on aqueous dispersions of polystyrene particles. Five measurements of 90s duration were taken for all autocorrelation, cross-correlation, and modulated cross-correlation data that were collected. DLS measurements were made on 100 nm nominal diameter (<15% polydispersity, Thermo) polystyrene particles in 0.1 mM NaCl in 5.0/4.2 mm outer/inner diameter cylindrical glass cuvettes. SLS measurements were performed on 430 nm diameter polystyrene (<3% polydispersity, Thermo) in $H_20$ placed in 10/9 mm outer/inner diameter glass cuvettes, except for dilute measurements which were made in quartz cuvettes of 10/8 mm outer/inner diameter (Helma Optik). Multiple scattering correction was applied to turbid SLS measurements by multiplying the intensity times the square root of the cross-correlation intercept measured for the turbid sample as a function of angle, divided by that measured for 100 nm polystyrene spheres at 0.002 % w/w in $H_20$. This correction factor gives the ratio of singly-scattered light to the total intensity measured for each scattering angle. DLS particle sizing and intercept determination were performed using a single exponential fit. Fitting of SLS data was achieved with a manual best fit using Mie theory calculations from freely available software (MiePlot, www.philiplaven.com/mieplot.htm).

**Results**

First, DLS measurements at a fixed scattering angle of 90-deg were made of a dilute sample of 100 nm polystyrene in $H_20$ (0.002 %w/w, 99.2% transmission) in autocorrelation, 3D cross-correlation, and then modulated 3D cross-correlation. Fig. 2(a) shows a comparison of correlation data measured using each of the three methods. An intercept of 0.971±0.005 close to the theoretical value of 1 is obtained for the autocorrelation function, demonstrating excellent coherence of the incident laser in the scattering volume and alignment of single-mode fibers to a single laser speckle in the far-field. For the standard 3D cross-correlation measurement we expect a four-fold reduction in the intercept due to cross-talk between the two beam-detector pairs as well as further losses due to the incomplete overlap of the incident beams and symmetrically, the focal volumes of the collection optics. The intercept for the cross-correlation technique is measured to be 0.188±0.002. Driving the AOMs and detection gating electronics in order to temporally separate the scattering data collected from the two beam-detector pairs leads to an intercept of 0.767±0.002, yielding essentially 100% of the expected 4x improvement in performance. The difference is slightly more than the anticipated four-fold improvement due to a very small compromise in the optical alignment when the AOMs are not activated.

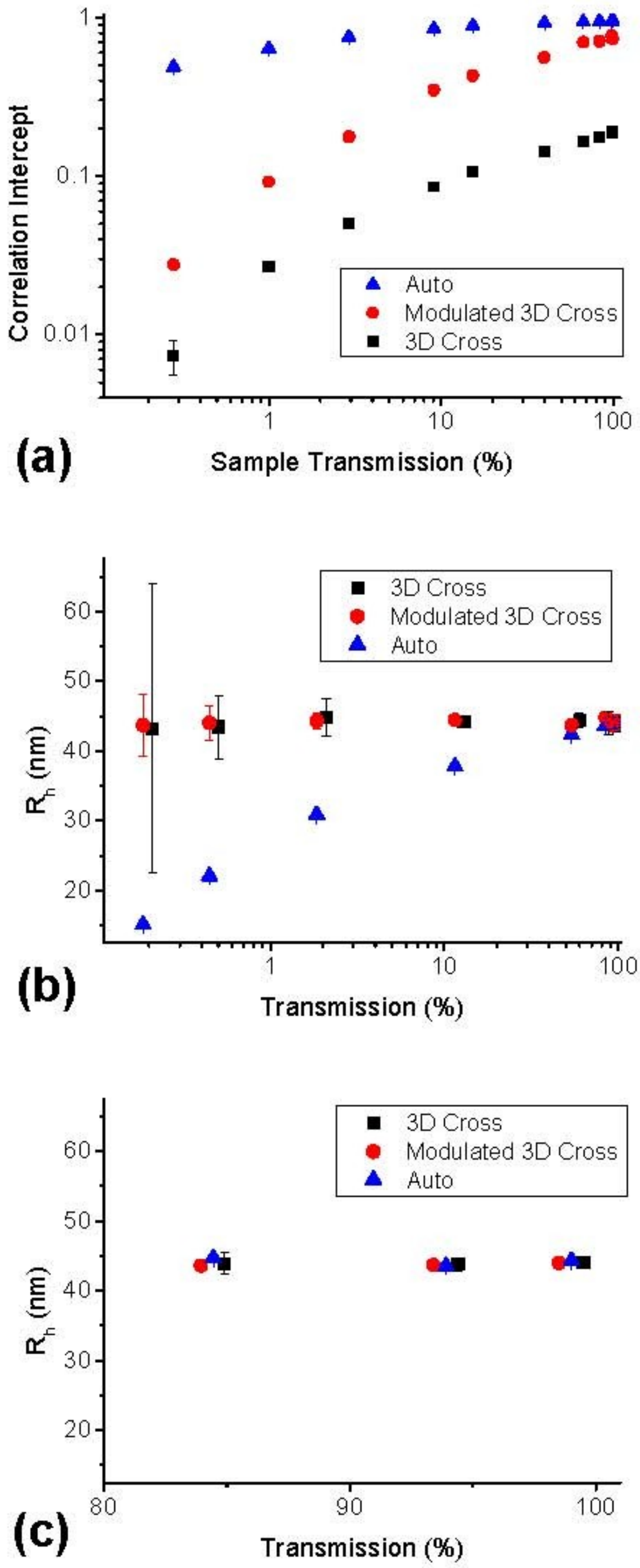

**FIG. 3.** (Color Online) Correlation function intercept **(a)** and extracted hydrodynamic radius **(b)** as a function of sample turbidity for each of the three correlation methods. Large errors for autocorrelation and improved precision of modulated over standard 3D cross-correlation are apparent. 3D cross measurements in (b) have been slightly shifted horizontally to ease comparison. **(c)** Linear re-plot of (b) illustrating excellent agreement of the three correlation methods for dilute samples.

The influence of the choice of modulation frequency on the intensity correlation function intercept is given in Fig. 2(b). Two data sets are shown, one for which a simple gating of the APDs is performed in unison with AOM activation, and a second in which more complex high-speed electronics are implemented for additionally blanking photon detector output during the rise and fall-times of the AOMs. A sharp decrease in the intercept is visible for the more basic gating method at modulation frequencies exceeding approximately 50 kHz, owing to the increased temporal overlap of the two scattering experiments due to finite rise and fall times of AOM operation. In contrast, the full four-fold improvement in the cross-correlation intercept is maintained upwards of 500 kHz for the gating scheme in which data is captured only once the AOMs and therefore the beams incident on the scattering cell have reached a steady state.

In order to demonstrate the utility of this modulated 3D cross-correlation technique for the characterization of turbid colloidal samples, DLS measurements were performed and analysed for the same system as described above as well as for increasingly more concentrated solutions up to 2.5% w/w (and correspondingly 0.2% optical transmission). Fig. 3(a) shows the evolution of the intercept as a function of optical transmission for autocorrelation, cross-correlation and modulated cross-correlation techniques. The relative magnitude of the cross-correlation intercept is reduced at high particle concentrations as this directly provides a measure of the contribution of singly-scattered light to the overall scattered signal. As the intercept drops, the signal begins to fall into the measurement noise and therefore the extraction and analysis of any information contained in this data having a reduced signal-to-noise ratio (SNR) becomes less reliable. The improved magnitude of the correlation intercept for the method introduced in this work leads to more precise DLS measurements for highly turbid samples, as demonstrated in Fig. 3(b). It can also be seen that autocorrelation measurements quickly lose accuracy as sample transmission drops below approximately 90%. As we should expect, for dilute samples the three methods show excellent agreement as seen in Fig. 3(c).

As for the case of time dependent measurements, static light scattering (SLS) measurements of scattered intensity as a function of scattering vector rely on the properties of singly-scattered light. Extraction of only the single-scattering contribution is therefore of great importance for the precise determination of particle form and structure factors for a turbid sample. It is known that SLS measurements of multiple-scattering systems can be improved by correcting the absolute scattering intensities by the amount of single scattering present as a function of scattering angle. The single scattering contribution is calculated as the square root of the ratio of the measured cross-correlation intercept to that achieved for a dilute sample with no multiple scattering at each scattering vector. SLS measurements were first made for turbid (0.01 % w/w) and dilute ($3\times10^{-5}$ % w/w) samples of 430 nm polystyrene particles in water. As shown in Figure 4(a), a Mie fit to the dilute data for a fixed particle refractive index of $n = 1.58$ gives a mean particle diameter of 210 nm with a polydispersity of 1%, in good agreement with measurements provided by the manufacturer. However, multiple scattering in the turbid sample strongly erodes the form factor minimum and the data deviate significantly from the expected response with an apparent polydispersity misrepresented by more than an order of magnitude. In order to suppress the strong effects of multiple scattering, the single-scattering correction was applied to the data of the turbid sample by measuring the cross-correlation intercept as a function of scattering vector. Figures 4(b) and (c) illustrate the results of implementing such a correction as measured with the modulated 3D cross-correlation and standard 3D cross-correlation techniques, respectively. While both methods recover the overall shape of the curve measured for the dilute sample, the greater SNR provided by the modulated 3D cross-correlation technique gives a more accurate and robust result. Precise corrections to the total scattered intensity are especially important in the form factor minimum where multiple scattering dominates and the cross-correlation intercept tends towards zero.

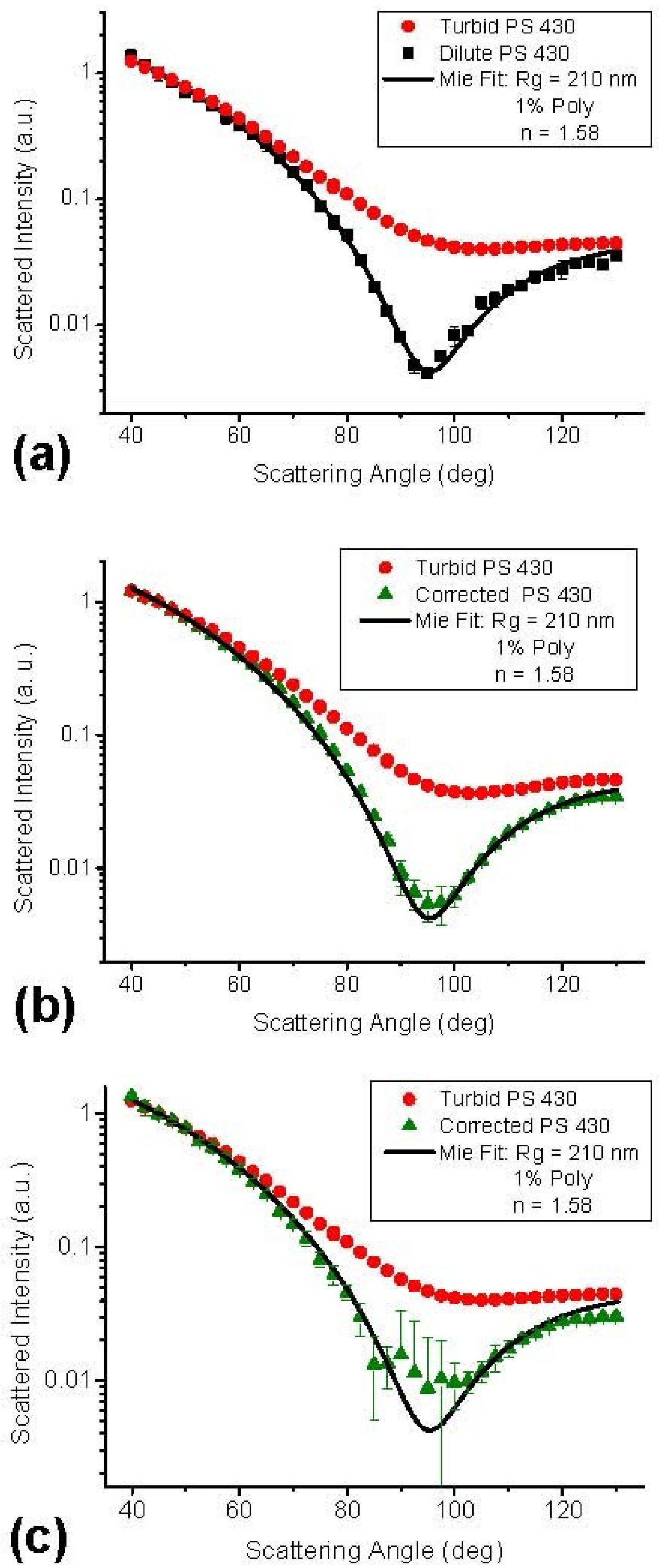

**FIG. 4.** (Color Online) **(a)** SLS measurements of turbid and dilute suspensions of 430 nm polystyrene particles, along with a Mie fit to the dilute data. Corrected SLS measurements of the turbid sample using modulated **(b)** and standard **(c)** 3D cross-correlation techniques.

**Discussion**

Temporally isolating the two beam-detector pairs in the 3D geometry significantly improves the SNR of static and dynamic light scattering measurements. This advantage can be used to reduce measurement times, yielding significantly improved temporal measurement resolution for systems displaying rapid changes in size, shape or interactions. In addition, the technique enables measurement of more highly scattering samples than previously possible with the traditional 3D setup. The use of small cylindrical cuvettes or of a square cuvette aligned in the $\theta$-$2\theta$ arrangement can additionally be adapted to this setup to further improve measurement statistics.

The hardware required for modulating the two beam-detector pairs was added to a commercial light scattering instrument without limiting any of the features of the existing tool. In addition, a stable optical alignment has been maintained over several months during which this work was performed. This is especially critical for the 3D cross-correlation arrangement in which alignment errors of 10 microns or less can have a significant impact on instrument performance. Furthermore, the modulation mode can be exchanged for standard 3D cross-correlation or traditional autocorrelation within seconds if dynamics at time scales faster than the modulation speed are of interest.

An additional configuration for the modulated 3D cross-correlation technique is possible wherein a single AOM can provide the dual functionality of generating each of the two incident beams and switching at high frequencies between them. Such a scheme eliminates the need for a beamsplitter that must be constructed with stringent alignment tolerances, requires only a single AOM and associated RF amplifier, and has the further benefit of increasing the illumination intensity by a factor of two. However the tradeoff is that one cannot use the instrument in the standard 3D scheme since two incident beams with stable intensities cannot be generated simultaneously. In this work we have opted against this configuration for a more straightforward comparison with the traditional 3D technique and to ease integration with, and ensure continued robust operation of an existing commercial instrument.

The magnitude of the cross-correlation intercept which directly impacts measurement performance is linked to the efficiency of the optical and electronic signal separation as well as to the response times of each of these modules. By adopting a detector gating scheme that ignores photon events during transient AOM response, the measured gain in intercept maintains the theoretical four-fold improvement even at modulation frequencies exceeding 500 kHz. AOMs were selected for this application as they are relatively inexpensive, require low driving voltages, and can preserve the shape, polarization, and propagation vector of the laser beam. Considering that the majority of highly turbid systems will display slow dynamics, these devices as implemented here possess sufficient modulation speed to accurately measure such samples. However, it is possible to improve the AOM response speed by approximately an order of magnitude by adopting either a free space or fiber-coupled arrangement in which the beam is focused through the AOM. Faster modulators using electro-optic crystals are also readily available, as are directly modulated laser diodes. Therefore, the potential to extend the technique to faster relaxing turbid systems certainly exists.

The major previous criticism of the 3D DLS technique was the reduced cross-correlation intercept. Using the method presented in this work, a light scattering instrument based on the 3D geometry no longer gives any compromises in regards to measurement times or performance when compared to a traditional single-beam autocorrelation light scattering instrument. In addition, it extends these methods to highly turbid and even opaque samples, eliminates the need for time consuming dilution and verification, and prevents user error due to inadequate dilution. Therefore, the modulated 3D cross-correlation technique presented here substantially improves the performance and flexibility of this instrument and will enable its applicability to an even wider range of samples.

**Acknowledgements**

We would like to thank Peter Schurtenberger and Charles Völker for fruitful discussions. We would also like to gratefully acknowledge financial support from the CTI (Innovation promotion agency of the Swiss Confederation) under grant 9884.1.